# Constraining the vertical distribution of coastal dust aerosol using OCO-2 O₂ A-band measurements

Zhao-Cheng Zeng[1,2], Sihe Chen[2,3], Vijay Natraj[4], Tianhao Le[2], Feng Xu[4], Aronne Merrelli[5], David Crisp[4], Stanley P. Sander[4], and Yuk L. Yung[2,4]

[1]University of California, Los Angeles, USA;

[2]California Institute of Technology, Pasadena, USA;

[3]National University of Singapore, Singapore;

[4]Jet Propulsion Laboratory, California Institute of Technology, Pasadena, USA;

[5]University of Wisconsin-Madison, Madison, Wisconsin, USA.

Correspondence to: Z.-C. Zeng (zcz@gps.caltech.edu)

Address: 1200 E. California Blvd MC-150-21, Pasadena, CA 91125

**Abstract:**

Quantifying the vertical distribution of atmospheric aerosols is crucial for estimating their impact on the Earth's energy budget and climate, improving forecast of air pollution in cities, and reducing biases in the retrieval of greenhouse gases (GHGs) from space. However, to date, passive remote sensing measurements have provided limited information about aerosol extinction profiles. In this study, we propose the use of a spectral sorting approach to constrain the aerosol vertical structure using spectra of reflected sunlight absorption within the molecular oxygen ($O_2$) A-band collected by the Orbiting Carbon Observatory-2 (OCO-2). The effectiveness of the approach is evaluated using spectra acquired over the western Sahara coast by comparing the aerosol profile retrievals with lidar measurements from the Cloud-Aerosol Lidar and Infrared Pathfinder Satellite Observation (CALIPSO) Cloud-Aerosol Lidar with Orthogonal Polarization (CALIOP). Using a radiative transfer model to simulate OCO-2 measurements, we found that high-resolution $O_2$ A-band measurements have high sensitivity to aerosol optical depth (AOD) and aerosol layer height (ALH). Retrieved estimates of AOD and ALH based on a look up table technique show good agreement with CALIPSO measurements, with correlation coefficients of 0.65 and 0.53, respectively. The strength of the proposed spectral sorting technique lies in its ability to identify spectral channels with high sensitivity to AOD and ALH and extract the associated information from the observed radiance in a straightforward manner. The proposed approach has the potential to enable future passive remote sensing missions to map the aerosol vertical distribution on a global scale.

**Highlights** (3-5 bullets; 85 characters including space)

- A spectral sorting approach is proposed for constraining aerosol profile from space;

- The approach is successfully applied to OCO-2 $O_2$ A-band measurements;

- Retrieved profile parameters show good agreement with CALIPSO measurements;

- The approach can be used to map the aerosol vertical distribution on a global scale.

## 1. Introduction

The vertical distribution of atmospheric aerosols plays an important role in regulating the Earth's energy budget by scattering and absorbing sunlight (direct effect) and via aerosol-cloud interactions (indirect effect; **IPCC, 2013; Zarzycki & Bond, 2010**). In addition, the detection of vertical distribution of aerosols, which contribute the largest environmental risk as air pollutants, enables assessment of their impact on public health (**Liu & Diner, 2017**). Third, aerosol scattering effects provide one of the most important sources of uncertainty in greenhouse gas (GHG) retrievals from space in the near infrared (**Kuang et al., 2002**). A better knowledge of the aerosol vertical distribution, which affects the light path length in the gas absorption channels, is required to achieve high accuracy GHG retrievals (**Crisp et al. 2008; Butz et al., 2009; O'Dell et al., 2018**). Satellite (e.g., MODIS and MISR; **Kahn et al., 2007**) and ground-based (e.g., AERONET; **Holben et al., 1998**) measurements have been accurately and continuously monitoring the global column total aerosol optical depth. However, to date, limited information on the vertical structure of atmospheric aerosols has been obtained from passive remote sensing measurements. Lidar measurements (e.g., CALIPSO; **Winker et al., 2010**) provide more information about the aerosol vertical distribution, but lidar instruments have a narrow swath and therefore require weeks to observe just a fraction of a percent of the planet's surface area; it is therefore very difficult to obtain global coverage. This necessitates a new era of research pursuing alternative approaches to solve this challenging problem.

The use of oxygen ($O_2$) absorption measurements to constrain cloud and aerosol profiles was first proposed by **Yamamoto & Wark (1961)**. The physical basis is that (1) $O_2$ is uniformly distributed in the atmosphere with a mixing ratio of ~0.20955, (2) its spectrally-dependent absorption cross sections are reasonably well known (Drouin et al., 2017), and (3) as aerosols and clouds scatter light back to space, they leave distinctive signatures in different parts of the observed $O_2$ spectra, which are associated with the column aerosol/cloud optical depth and vertical structure. By studying these signatures, we can quantify the vertical distribution of aerosols and clouds. This technique has been successfully applied to quantify the cloud top pressure and cloud thickness (e.g., **O'Brien & Mitchell, 1992; Heidinger & Stephens, 2000;**

**Richardson et al., 2017**), quantify the impact of aerosol scattering on dry air mass used in $CO_2$ and $CH_4$ retrievals ( Bösch et al., 2006; 2011; Geddes and Bösch, 2015; O'Dell et al., 2018; Wu et al., 2018) and to characterize the impact of the 3-D structure of clouds on $CO_2$ retrievals (Massie et al., 2017). However, with respect to constraining the atmospheric aerosol vertical distribution, the majority of studies are still in the theoretical phase (e.g., **Hollstein & Fischer, 2014; Sanders et al., 2015; Colosimo et al., 2016; Ding et al., 2016; Davis et al., 2017; Hou et al., 2017**), with very few applications using real remote sensing measurements (e.g., **Dubuisson et al., 2009; Sanghavi et al., 2012; Xu et al., 2017; Zeng et al., 2018**).

Here we propose the use of a spectral sorting approach (**Liou, 2002; Richardson et al., 2017; Zeng et al., 2018**) for retrieving the total aerosol optical depth (AOD) and aerosol layer height (ALH) using hyperspectral $O_2$ A-band measurements from Orbiting Carbon Observatory-2 (**OCO-2; Crisp et al., 2008**), a NASA mission dedicated to measuring the concentration of atmospheric carbon dioxide ($CO_2$). This approach has been successfully applied to $O_2$ absorption measurements from the California Laboratory of Atmospheric Remote Sensing (CLARS; **Zeng et al., 2017; He et al., 2019**), a mountain-top Fourier Transform Spectrometer (FTS) overlooking the Los Angeles megacity, to detect the aerosol loading in the boundary layer. In this study, the effectiveness of this approach for satellite-based measurements is evaluated, using OCO-2 as a test case. We selected the western Sahara coast as the study area to minimize effects from variable land surface reflectance and to maximize the variability of aerosol layer height due to frequent dust storms over this coastal region. The development of an aerosol profile retrieval technique for passive remote sensing observations will be useful for future missions to produce maps of aerosol vertical structure on a global scale. In addition, the aerosol vertical structure constrained by this technique may provide a more efficient method for minimizing aerosol-related biases in $CO_2$ and $CH_4$ retrievals.

In **Section 2**, we describe how we obtain collocated measurements from OCO-2 and CALIPSO over the western Sahara coast. **Section 3** illustrates the sensitivity of the $O_2$ A-band to aerosol vertical structure using simulations of OCO-2 measurements. In **Section 4**, the AOD and ALH are retrieved using look-up tables (LUTs). The results are discussed in **Section 5**, and conclusions provided in **Section 6**.

## 2. Measurements of dust storms over the coastline in the western Sahara Desert

### 2.1 Western Sahara coast

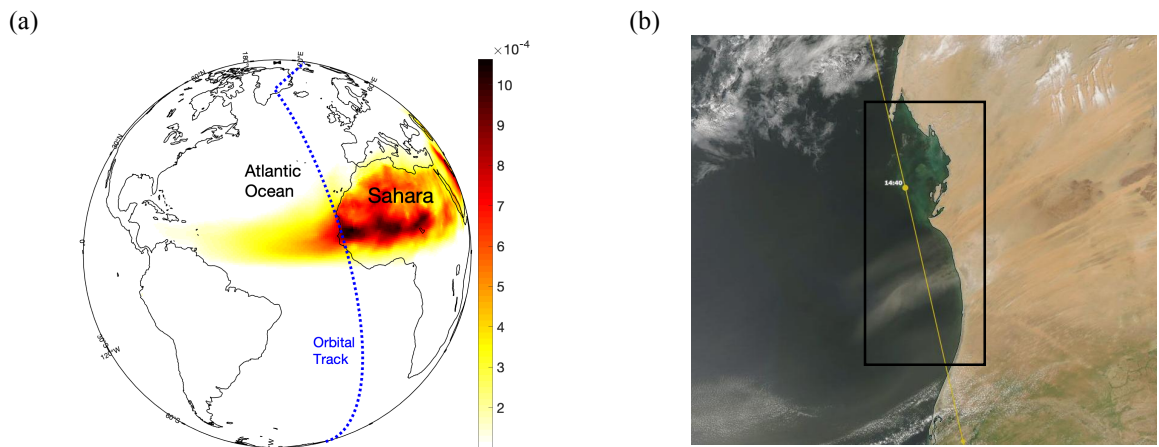

**Figure 1**. (a) Dust column density (kg/m$^2$) over the Mauritania coastline in the western Sahara Desert, obtained from the MERRAero 3-hour averaged monthly mean reanalysis dataset (**Buchard et al., 2016**) for May 2015. The sounding track (in blue) of both OCO-2 and CALIPSO overpassing the coastal region is also shown; (2) The Mauritania coastline in the western Sahara from MODIS/Aqua corrected reflectance (true color) on Sept 15, 2015. The sounding track (in yellow) of both OCO-2 and CALIPSO along the coast is also shown. The rectangle defines the study region between 17.5°N and 20.5°N used in this study.

The coastline of Mauritania in the western Sahara Desert faces the Atlantic Ocean (**Figure 1**). Each year, dust storms bring in hundreds of millions of tons of dust from the desert to the Atlantic Ocean (**Prospero and Mayo-Bracero, 2013**). Some of the dust reaches North and South America and affects the local air quality as well as climate, soil fertility, marine biology at large scales. A better knowledge of the dust vertical structure will improve our understanding of dust transport and assessment of its impact on global climate and local air quality. The large variability in dust vertical structure (see statistics from CALIPSO measurements in **Appendix A1**) and well characterized ocean surface reflectance make this coastal region an optimal testbed for remote sensing algorithms (**Dubuisson et al., 2009; Xu et al., 2017**). OCO-2 and CALIPSO also have regular overpasses over this coast, as shown in **Figure 1**, roughly once in every 16-day orbiting period. On roughly have of these opportunities between 15 May 2015 and 13 September 2018, the OCO-2 and CALIPSO observations were nearly bore-sighted. The sounding track between 17.5°N to 20.5°N is chosen here. The fact that the albedo of the ocean surface is small facilitates aerosol detection since the observed radiance tends to be highly sensitive to changes in aerosol loading. The Bidirectional Reflectance Distribution Function (BRDF) of the ocean surface is approximated by the Cox-Munk model (**Cox and Munk, 1954**) with wind speeds adopted from GEOS5 reanalysis data (see **Section 3.1**).

## 2.2 OCO-2 and CALIPSO measurements

The OCO-2 and CALIPSO satellites fly in the A-Train constellation (**L'Ecuyer and Jiang, 2010**). At an altitude of about 705 km, the A-Train follows a sun synchronous orbit with a ground track repeat cycle of 16 days and an equator overpass time around 13:30 hours local time. Between 15 May 2015 and 13 September 2018, the OCO-2 and CALIPSO navigation teams collaborated to align the ground tracks of the two satellites (**Figure 2(a)**); therefore, their measurements are collocated when OCO-2 is acquiring observations near the local nadir, which is does on roughly half of all orbits.

OCO-2 was launched in July 2014 to measure the abundance of atmospheric $CO_2$, one of the most important greenhouse gases (**Crisp et al., 2008; Eldering et al., 2017**). Using imaging grating spectrometers, it collects high-resolution spectra of reflected sunlight in the $O_2$ A-band centered at ~0.765 μm, the weak $CO_2$ band centered at 1.61μm, and the strong $CO_2$ band centered at 2.06 μm. The spectral resolution in the $O_2$ A-band, is ~0.04 nm (**Figure 2(b)**). The $O_2$ A-band spectra have high sensitivity to changes in the light path. They are primarily used to determine the surface pressure, to screen for optically thick clouds (**Taylor et al., 2016**) and to quantify the total optical depth (AOD) and vertical distribution of aerosols for soundings that are sufficiently optically thin (AOD < 0.3) to yield accurate estimates of the column average $CO_2$ dry air mole fraction, XCO2 (O'Dell et al., 2018).

The OCO-2 spectrometers collect 8 spectra at 0.333 second intervals across a narrow (0.1°) swath, yielding footprints with a spatial resolution of less than 3 $km^2$. In this study, we use the OCO-2 footprint that is the closest to the corresponding CALIPSO sounding. The OCO-2 Level 1B (L1B) calibrated, geolocated science spectra, retrospective processing version V8r (OCO2_L1B_Science) are collected from NASA's Earth Observing System Data and Information System (**NASA, 2019c**). Over the ocean, OCO-2 has two observation modes: the "nadir" mode, where the instrument collects data while pointed at the local nadir, and the "glint" mode, where the instrument is pointed at a point near the glint spot, where sunlight is specularly reflected from the Earth's surface. In this study, we focus on "nadir" mode measurements, which are more closely aligned with the CALIPSO observations and have less contribution from the ocean surface. Over the western Sahara coast, there are a total of 45 nadir sounding tracks from 2014 to 2018; excluding tracks with dense clouds and without collocated CALIPSO measurements, there are a total of 27 sounding tracks available, which are listed in **Appendix A2**.

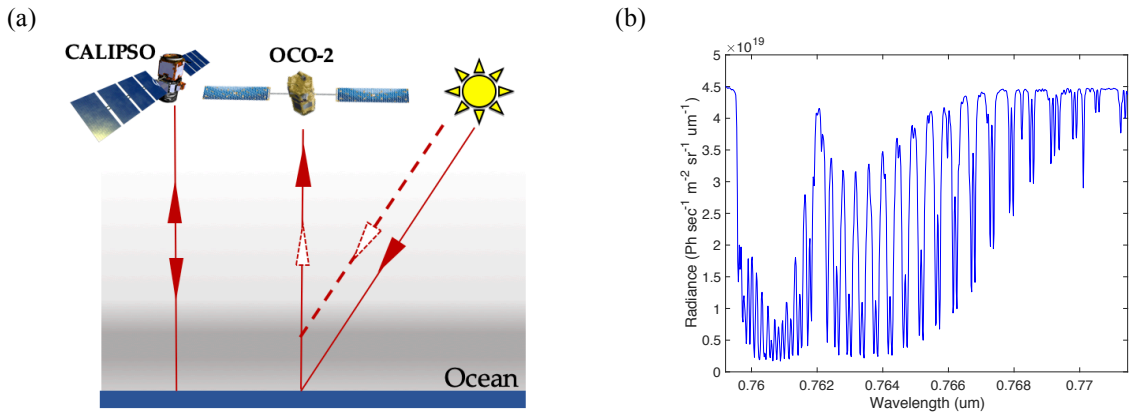

**Figure 2**. (a) Illustration of OCO-2 and CALIPSO observations in the A-Train constellation. CALIPSO is an active instrument while OCO-2 employs passive remote sensing. The aerosol layer in the atmosphere and light scattering due to aerosols are also indicated; (b) $O_2$ A-band radiance simulated by the OCO-2 forward radiative transfer (RT) model over the western Sahara coast. The aerosol vertical profile (**Equation (1)**) is defined by the mean values of AOD and ALH derived from CALIPSO (see **Appendix A1**). The meteorology and observation geometries are adopted from OCO-2 soundings on 15 September 2015.

CALIPSO was launched in 2016. Prior to September 2018, it flew in the A-Train about 7 minutes behind OCO-2. Its primary instrument, CALIOP, uses an active lidar, which collects data near the local nadir at two-wavelengths (532nm and 1064 nm) to probe the vertical structure and properties of clouds and aerosols (**Winker et al., 2010**). The standard data product from CALIPSO used in this study is the Level-2 5-km aerosol profile (CAL_LID_L2_05kmAPro-Standard-V4-20; **NASA, 2019b**). The aerosol profile products are reported at a uniform spatial resolution of 60 m vertically and 5 km horizontally. This dataset includes (1) column AODs at both 532 nm and 1064 nm; (2) column cloud optical depth (COD) at 532 nm; (3) column stratospheric AOD at both 532 nm and 1064 nm; and (4) the profile of aerosol extinction coefficient at 532 nm. Here, we estimated the column AOD at 765 nm in the $O_2$ A-band by interpolation using the Ångström exponent law (**Seinfeld and Pandis, 2016; Appendix A3**). Column COD is used to filter data with cloud contamination. The extinction coefficient profile is use to derive ALH (see **Appendix A4**). An illustration of the extinction coefficient profile, column AOD and COD, and the derived ALH are shown in **Figure 3**; the corresponding $O_2$ A-band continuum level radiance, which is related to the column AOD, is also shown.

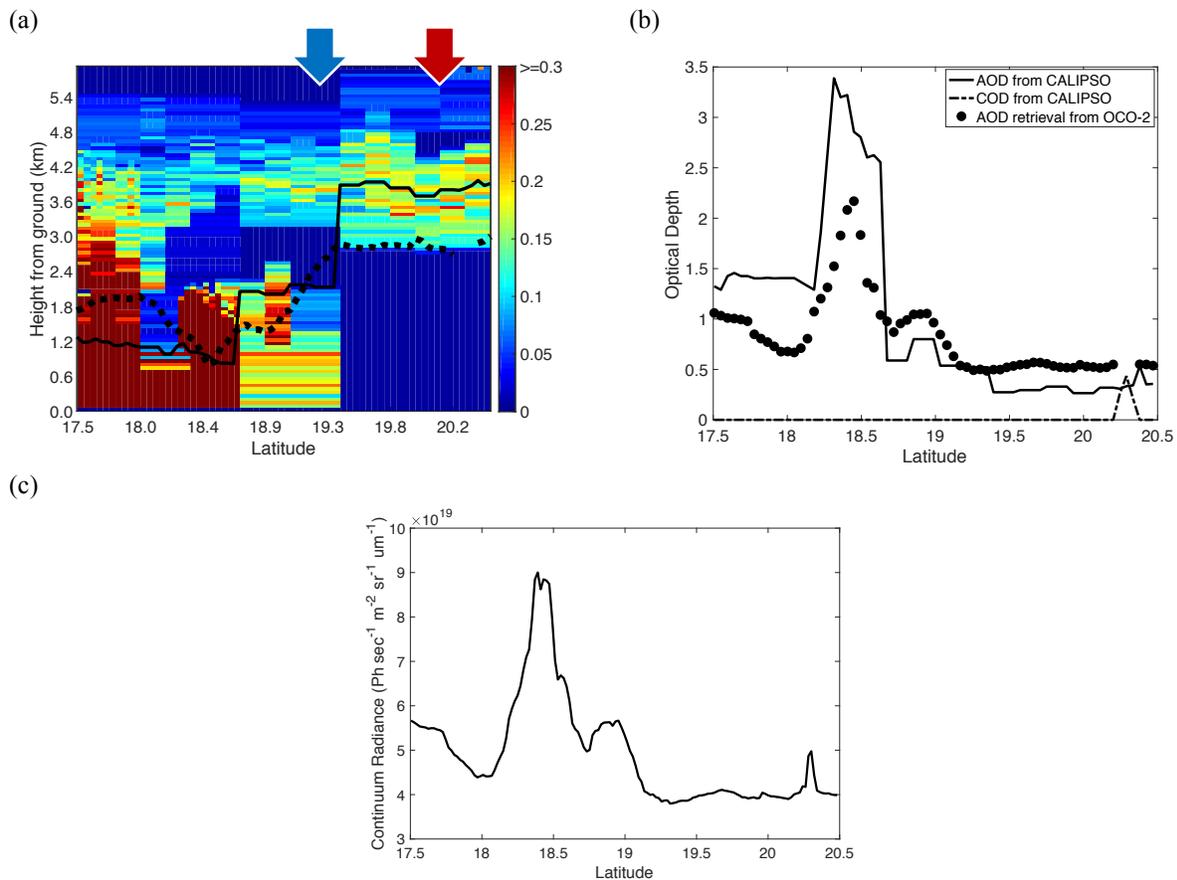

**Figure 3**. (a) Example of an aerosol extinction coefficient profile at 532 nm from CALIPSO along the sounding track across the western Sahara coast. The measurements were made on 15 September 2015. The estimated ALH (solid black line) from CALIPSO and the retrieved ALH from OCO-2 (dashed black line; see **Section 4**) are also shown. The two arrows at 19.17°N (blue) and 20.02°N (red) indicate two examples of soundings that are described in **Section 3.3**; (b) AOD and COD retrieved from CALIPSO measurements and the estimated AOD (see **Section 4**) by the algorithm developed in this study from OCO-2 measurements; (c) Corresponding OCO-2 $O_2$ A-band continuum level radiance over the same sounding track.

## 3. Sensitivity of $O_2$ A-band measurement to aerosol vertical structure
### 3.1 OCO-2 forward model

The Atmospheric $CO_2$ Observations from Space (ACOS) retrieval algorithm used to produce the OCO-2 Level 2 Full Physics (FP) product incorporates a forward radiative transfer (RT) model that simulates the observed radiance in the $O_2$ A-band as well as the $CO_2$ bands at 1.61 and 2.06 µm (**O'Dell et al., 2018; Merrelli et al., 2015**). The OCO-2 forward RT model combines the LInearized Discrete Ordinate RT (LIDORT) model (**Spurr, 2002**) to solve the scalar RT equation with a two orders of scattering (2OS) technique (**Natraj and Spurr, 2007**) to account for polarization in a scattering atmosphere. The OCO-2 forward model has been fully tested and validated by comparing against VLIDORT (**Spurr, 2006**), a linearized pseudo-spherical vector discrete ordinate RT model, for the nadir, glint, and target observing modes of OCO-2. The OCO-2 forward model generates the radiance using multiple inputs, including (1) spectrally resolved surface and atmospheric optical properties (aerosol optical properties and vertical distributions, gas absorption and scattering cross-sections, and surface type); (2) solar model and instrument Doppler shift; (3) instrument model, such as instrument line shape (ILS) and polarization response; and (4) satellite observing geometries. A detailed description of the OCO-2 forward model can be found in the OCO-2 Level-2 FP Retrieval Algorithm Theoretical Basis Document (**NASA, 2014**) and in the literature (Bösch et al., 2011; Crisp et al., 2012; O'Dell 2012; 2018). The OCO-2 forward model is publicly available at **NASA-JPL (2017)**.

In this study, the forward RT model for the OCO-2 FP algorithm Build v8 is used. The *a priori* wind speed for the Cox-Munk model and the atmospheric profiles (from 1 hPa to the ocean surface) for the RT calculation are selected from the GEOS5 FP-IT reanalysis (**Lucchesi et al., 2013**). Atmospheric molecular absorption cross-sections are obtained from the ABSCO database Version 5 developed for OCO-2 (**Drouin et al., 2017**). The two-year MERRA climatology (2009-2010; **Rienecker et al., 2011**) indicates that dust is the dominant aerosol (with a fraction close to 1.0; **NASA, 2014**) in terms of averaged AOD in the $O_2$ A-band over the western Sahara coast. Therefore, we assume one type of aerosol (dust) in the OCO-2 forward model; the relevant optical properties, including single scattering albedo and phase function, are described in **Appendix A5**.

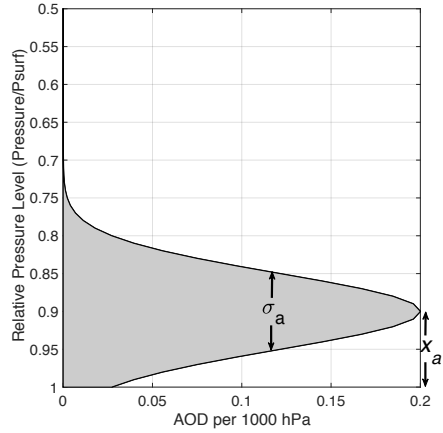

**Figure 4.** Gaussian profile shape used to approximate the vertical distribution of aerosol in the OCO-2 forward model. As described in **Equation (1)**, there are three parameters to define the shape: column AOD characterizes the total aerosol amount; peak height ($x_a$) and aerosol layer width ($\sigma_a$) characterize the vertical distribution. The vertical coordinate ($x$) is defined using relative pressure $P/Psurf$, where $P$ is the pressure in a specific layer and $Psurf$ is the surface pressure.

## 3.2 Sensitivity of radiance to aerosol vertical structure

As in **O'Dell et al. (2018)**, a Gaussian shape is used to approximate the vertical distribution of dust aerosol over the western Sahara coast, as shown in **Figure 4**. The Gaussian profile shape $S(x)$ is given by:

$$S(x) = C * \exp\left(-\frac{(x - x_a)^2}{2\sigma_a^2}\right) \tag{1}$$

where the vertical coordinate ($x$) is defined using relative pressure $P/Psurf$; $P$ is the pressure in a specific layer and $Psurf$ is the surface pressure. By definition, $x$ ranges between zero and one, corresponding to levels at the top of the atmosphere and at the surface, respectively. $x_a$ is the peak height and $\sigma_a$ is the one standard deviation ($\sigma$) width of the dust aerosol layer. $C$ is a normalization factor such that the integral of $S(x)$ equals the total AOD. The mass center of the CALIPSO aerosol layer is quantified by ALH. The methodology for conversion between ALH from CALIPSO and the peak height ($x_a$) in the OCO-2 forward model is described in **Appendix A4**. *A priori* values for $x_a$ and $\sigma_a$ are 0.9±0.2 and 0.05±0.01, respectively.

To quantify the sensitivity of the $O_2$ A-band to changes in AOD, ALH, and aerosol layer width (ALW), we execute the OCO-2 forward model to first simulate the radiance for different sets of aerosol profiles and then calculate the sensitivity of the radiance to changes in AOD, ALH, and ALW, respectively. The AOD and ALH are the means calculated from all available CALIPSO data along the western Sahara coast (**Appendix A1**). A sample $O_2$ A-band radiance spectrum obtained using this procedure is shown in **Figure 2(b)**. We use finite difference (assuming a perturbation of 1%) to estimate the sensitivity to the three aerosol parameters. As shown in **Figure 5(a)**, the derivatives of the radiance with respect to AOD are highly correlated with the observed radiance spectrum, especially in the continuum absorption regions. This suggests that when the aerosol is bright, the surface albedo is low and uniform across the spectral region and the viewing and illumination geometry are invariant, the continuum radiance is approximately proportionate to the AOD along the light path. Therefore, the continuum radiance provides a constraint on AOD when the surface and aerosol have substantially different optical properties. The ALH sensitivity, however, has peaks in intermediate absorption lines and troughs in continuum channels and strong absorption lines. This is because photons scattered by aerosols located higher up in the atmosphere will undergo shorter or longer absorption paths and therefore different $O_2$ absorption depths than those reflected only by the surface when the surface is relatively dark. The continuum level has little change because the absorption by $O_2$ is negligible there since scattering far outweighs absorption at $O_2$ A-band. Strong absorption lines show little change because they are saturated. As a result, intermediate absorption channels have the highest sensitivity. The $O_2$ A-band is not sensitive to ALW, which is consistent with the results of **Butz et al. (2009)**. These patterns in the sensitivity can be clearly observed when we sort the spectra in order of increasing radiance (**Figure 5(b)**). It is evident that the continuum level has the largest response to AOD while the intermediate absorption channels (sorted channel index between 150 and 250) have the largest response to ALH. In other words, the $O_2$ A-band contains at least two pieces of information for constraining the aerosol vertical structure: the continuum level provides constraints on the total AOD and the intermediate absorption channels provide constraints on the ALH.

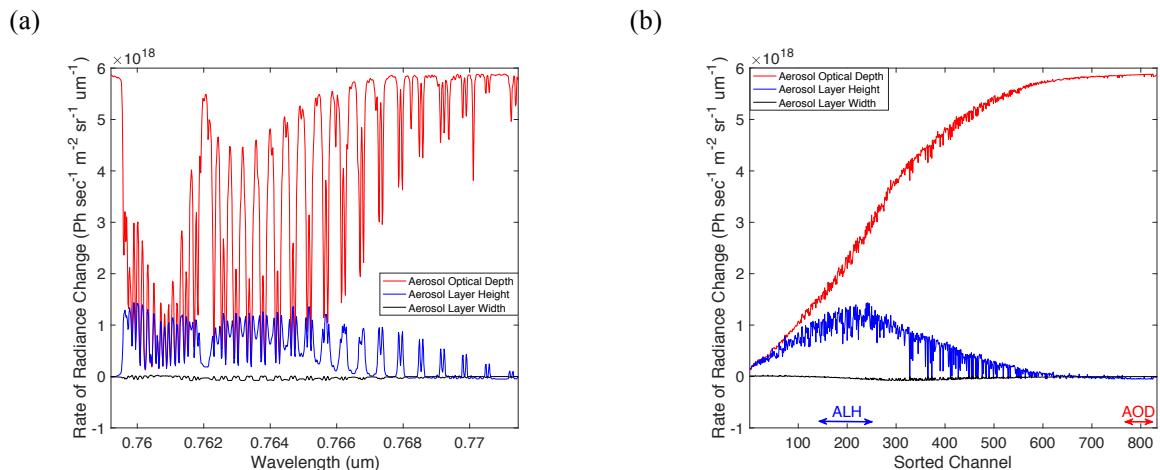

**Figure 5**. (a) Sensitivity of OCO-2 measured $O_2$ A-band radiance to the three parameters (AOD, ALH and ALW) that define the vertical structure of aerosols. The sensitivity for each of the three parameters is calculated by finite difference assuming a perturbation of 1%; (b) Same as (a) but the sensitivities for all channels have been sorted in order of increasing radiance. The channels used for retrievals of ALH (blue arrow) and AOD (red arrow) are also indicated.

### 3.3 Comparison of simulations and measurements

The performance of the OCO-2 forward model is evaluated through comparisons with measurements from OCO-2. The focus here is to investigate whether the forward RT model can reproduce the enhanced radiance in the intermediate absorption channels due to increases in ALH. We specifically selected two soundings on the same OCO-2 orbit track with similar AOD (i.e., nearly identical continuum radiance level) but different ALH (calculated from CALIPSO). The locations of the two soundings are indicated by the red (high aerosol layer) and blue (low aerosol layer) arrows in **Figure 3(a)**. The measurements are shown in the upper panel of **Figure 6**. The two soundings have spectra with similar continuum radiance levels (**Figure 6(a)**). We then sort the spectra in order of increasing radiance, as obtained for a simulation with ALH set to zero (black lines in **Figure 6(b)**, referred to as baseline hereafter). The sorted spectra are then normalized with respect to the continuum level. **Figure 6(b)** reveals differences in the intermediate absorption channels. The small wiggles in the sorted radiance are products of the variable wavelength dependence of oxygen absorption at different altitudes (Zeng et al., 2018). The differences are more evident when we calculate enhancements relative to the baseline, as shown in **Figure 6(c)**. The high (retrieved ALH=2.90 km and AOD=0.53) and low (retrieved ALH=2.38 km and AOD=0.53) aerosol layers lead to enhancements of about 5.5% and 4.2% on average, respectively, in the intermediate absorption channels. These features can be reproduced by the OCO-2 forward model, as shown in **Figures 6(d)**, **(e)**, and **(f)**, except that the measurements are noisier than the simulations. The comparisons (1) demonstrate the ability of the OCO-2 forward model to capture the response of the $O_2$ A-band to changes in the aerosol vertical profile, and (2) show the high sensitivity of OCO-2 measurements to aerosols.

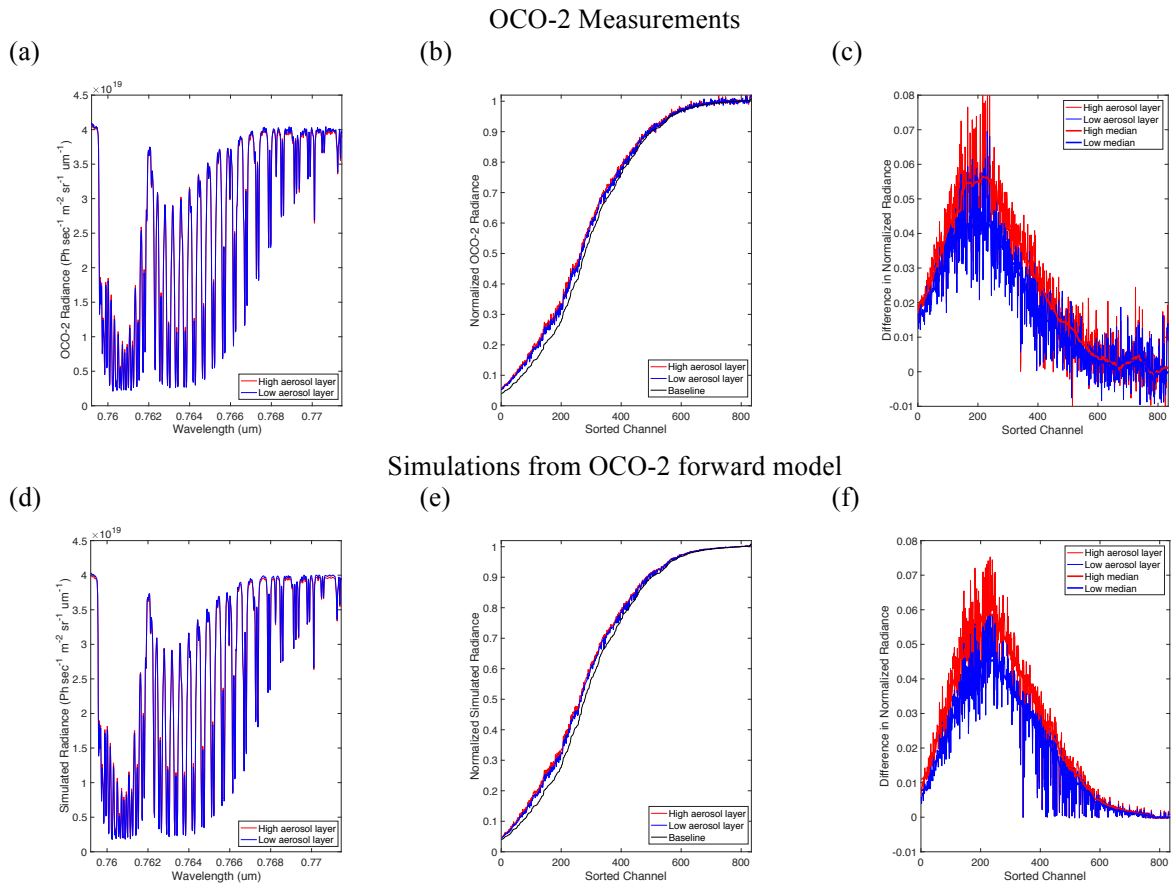

**Figure 6**. (a) Examples of measured radiance from two soundings on the same OCO-2 orbit track with similar AOD (i.e., identical continuum level radiance) but different ALH (as calculated from CALIPSO). The locations of the two soundings are indicated by the red (high aerosol layer with retrieved ALH=2.90 km and AOD=0.53) and blue (low aerosol layer with retrieved ALH=2.38 km and AOD=0.53) arrows in **Figure 3(a)**; (b) Same as (a) but sorted in order of increasing radiance, as obtained for a simulation with ALH=0 (black line, hereafter referred to as baseline). The sorted spectra are normalized with respect to the continuum level; (c) Radiance enhancements relative to the baseline; (d), (e), and (f) are the same as (a), (b), and (c), respectively, but showing results from the OCO-2 forward model.

## 4. A spectral sorting approach for constraining aerosol profile

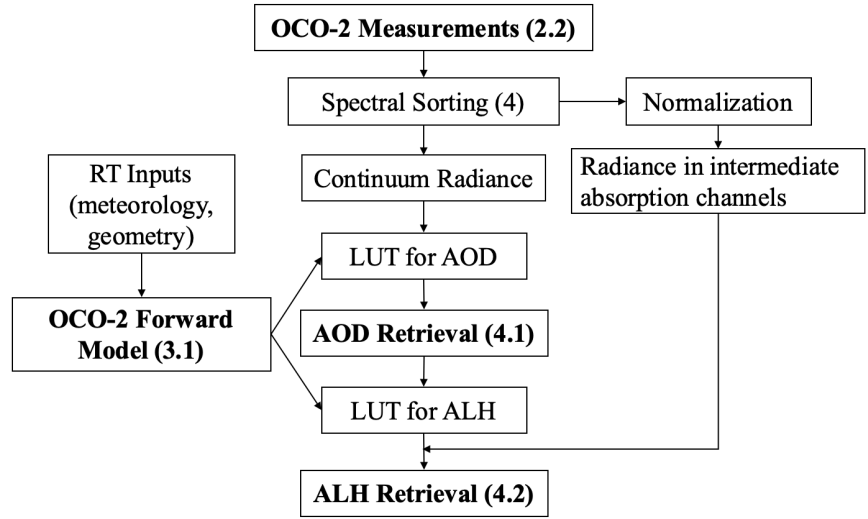

**Figure 7.** Schematic workflow for retrieving AOD and ALH from OCO-2 $O_2$ A-band measurements using a spectral sorting approach based on look up tables (LUTs). The numbers in parenthesis indicate the sections relevant to those topics.

A spectral sorting approach has been successfully applied to $O_2$ $^1\Delta$ band measurements at 1.27 μm from the CLARS-FTS instrument to profile aerosols within the planetary boundary layer in the Los Angeles megacity (**Zeng et al., 2018**). As shown in **Figure 7**, the retrieval approach is implemented by first constructing look up tables (LUTs) with different AODs and ALHs using the OCO-2 forward model. Due to changes in observation geometry and atmospheric conditions on a daily basis, the LUTs were built as a function of solar zenith angle (SZA), wind speed, atmospheric pressure, AOD, and ALH. The sensor zenith angle is not considered here because only nadir observing geometry is considered here. Examples of the LUTs are shown in **Appendix A6**. When implementing retrieval for a specific sounding, the corresponding LUT is selected according to the SZA, wind speed, and atmospheric pressure of the sounding. **Figures 8(a)** and **9(a)** show examples of the selected LUTs for the 27 days used in this study. As expected, the LUTs change with season due to the large variability in ocean surface reflectance associated with changing solar and illumination geometries. As described in **Section 3.2**, LUTs for AOD retrievals are built using the continuum level radiance in the $O_2$ A-band, while those for ALH retrievals are built using the radiance in selected intermediate absorption channels. In practice, the average radiance in the sorted channels between indices 780 and 810 is used for AOD retrieval. The average radiance in the sorted channels between indices 150 and 250 is used for ALH retrieval (**Figure 5(b)**).

The uncertainty in AOD and ALH retrievals primarily comes from uncertainties in the associated LUTs, which is related to uncertainties in the OCO-2 RT forward model and the associated model inputs. While it can be difficult to quantify LUT uncertainty analytically, several variables are important, including wind speed, atmospheric pressure, and aerosol optical properties. In this study, we estimate the uncertainty of LUTs using error propagation. Jacobians of $O_2$ A-band radiance to wind speed ($\frac{\partial rad}{\partial v}$), atmospheric pressure ($\frac{\partial rad}{\partial p}$), and aerosol single scattering albedo ($\frac{\partial rad}{\partial SSA}$), are estimated by finite difference using OCO-2 forward model. The uncertainty of a LUT is given by,

$$\delta_{LUT} = \sqrt{(\frac{\partial rad}{\partial v})^2 \delta_v^2 + (\frac{\partial rad}{\partial p})^2 \delta_p^2 + (\frac{\partial rad}{\partial SSA})^2 \delta_{SSA}^2} \tag{2}$$

in which $\delta_v$, $\delta_p$, $\delta_{SSA}$ are the uncertainties of wind speed (1.97 m/s), atmospheric pressure (250 Pa), and SSA (0.02), respectively. The perturbation amplitude, $\delta_v$ and $\delta_p$, are set at one sigma from GEOS FP-IT reanalysis data in the study region, while $\delta_{SSA}$ is estimated as one sigma from SSA retrievals from AERONET at Capo Verde, which is located by the Western Sahara coast. An example of the LUT uncertainty is shown in **Appendix A7**.

## 4.1 Retrievals of AOD

**Figure 8(a)** shows LUTs for all 27 days, color coded by month. To the first order, the continuum level radiance increases with increasing AOD due to stronger scattering by bright aerosols over dark surfaces. The difference in the continuum level radiance between different months is larger for low AODs. This is because when AOD is low, the surface reflectance dominates the contribution to the observed continuum level radiance. On the other hand, aerosol scattering dominates when AOD is large; therefore, differences between different months are small. Overlaying OCO-2 observations (AODs are derived from collocated CALIPSO measurements) with the LUTs shows that the two match very well, indicating the high accuracy of the OCO-2 forward model in capturing the seasonal variability of the observed spectra. The outliers (where observations and LUTs have large mismatch) may be due to the uncertainty in CALIPSO AOD retrievals, as discussed in **Omar et al. (2013)** and **Appendix A8**.

We retrieve AOD by linearly interpolating the continuum level radiance based on the LUTs. About 87% of the measurements fall into the calculated LUT range and produce valid AOD retrievals. The comparison between AOD retrievals and collocated AOD retrievals from CALIPSO is shown in **Figure 8(b)**. They show a high correlation, with an $R^2$ value of 0.65. The root mean square error (RMSE) is 0.29. On average, the estimation error is about 0.65% for each sounding. The uncertainty of estimated AOD, indicated by error bars in **Figure 8(b)**, is quantified by multiplying the Jacobian of estimated AOD to LUT and the LUT uncertainty estimated from Equation (2).

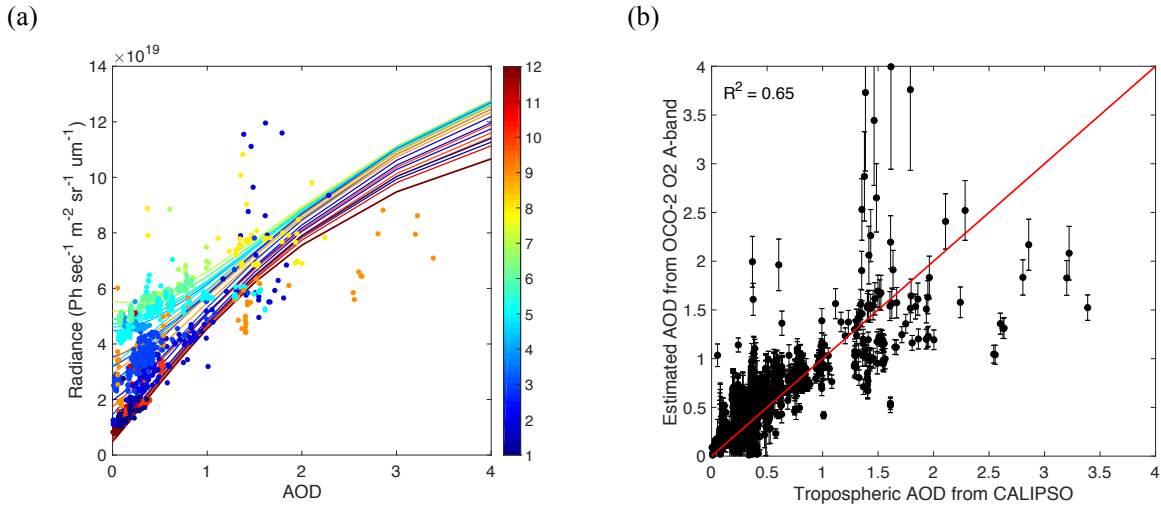

**Figure 8**. (a) Examples of AOD look up tables (LUT; solid lines) averaged for each of the 27 days (one line for each day), constructed using the OCO-2 forward model, color coded by month and overlaid with OCO-2 $O_2$ A-band continuum level radiance measurements (dots). These LUTs are selected from the LUT database (as shown in **Figure A6**) according to the SZA, wind speed, and atmospheric pressure of the sounding. AOD values for the OCO-2 observations are adopted from collocated CALIPSO measurements; (b) Scatter plots of AOD retrievals from OCO-2 $O_2$ A-band and collocated CALIPSO measurements. About 87% of the measurements fall into the calculated LUT range and produce valid AOD retrievals. The root mean square error (RMSE) is 0.29 for AOD retrievals. The error bars indicate the retrieval uncertainty. Some of the error bars are outside the figure box.

## 4.2 Retrievals of ALH

**Figure 9(a)** shows examples of LUTs for ALH retrieval for four seasons. We can see that, even though all of them show an increasing trend in the radiance in intermediate absorption channels as AOD increases, the behavior is very different under different AOD conditions. As expected, the change is smaller when AOD is low, due to weaker aerosol scattering. In January and October, the ocean surface is very dark because the solar zenith angle is large and far away from the ocean "glint" angle over the western Sahara coast. Therefore, when AOD is low, Rayleigh scattering dominates. The mean height of these atmospheric scatterers is determined by their density profiles. If the ocean surface is completely dark, the mass center of the atmosphere is about 6.3 km (assuming the scale height of the atmosphere is about 9 km). Therefore, the normalized radiance in the intermediate absorption channels is higher in October and January when AOD is small than when AOD is large. In April and July, however, the solar zenith angle is smaller and therefore the ocean surface reflectance is higher. As a result, the ocean surface reflectance dominates the contribution to the observed radiance. Therefore, the normalized radiance in the intermediate absorption channels increases as AOD increases, keeping ALH fixed.

We retrieve ALH by linearly interpolating the normalized radiance in the intermediate absorption channels (and using the AOD as retrieved in **Section 4.1**), based on the LUTs. Uncertainties in ALH retrievals are estimated as a function of the AOD uncertainty. We use the upper and lower bounds of the AOD retrieval to quantify the upper and lower bounds of the ALH estimate, respectively. **Figure 9(b)** shows a comparison of ALH retrieval results with retrievals from CALIPSO. Again, there is a good correlation ($R^2 = 0.52$), with a RMSE of 0.68 km. On average, the estimation error is about 30% for each sounding. The uncertainty of ALH estimates, indicated by error bars in **Figure 9(b)**, is quantified by error propagation from the uncertainty of estimated AOD (**Figure 8(b)**), which, in turn, results from the LUT uncertainty. The scatter plot is color coded by the corresponding AOD. We find that the performance of the retrievals does not rely on the AOD, which is different from the findings of **Zeng et al. (2018),** where observations over land were used. An example of ALH and AOD retrievals along the sounding track on 15

September 2015 is shown in **Figure 3**. The OCO-2 retrievals using the spectral sorting approach clearly captures the general trends in AOD and ALH.

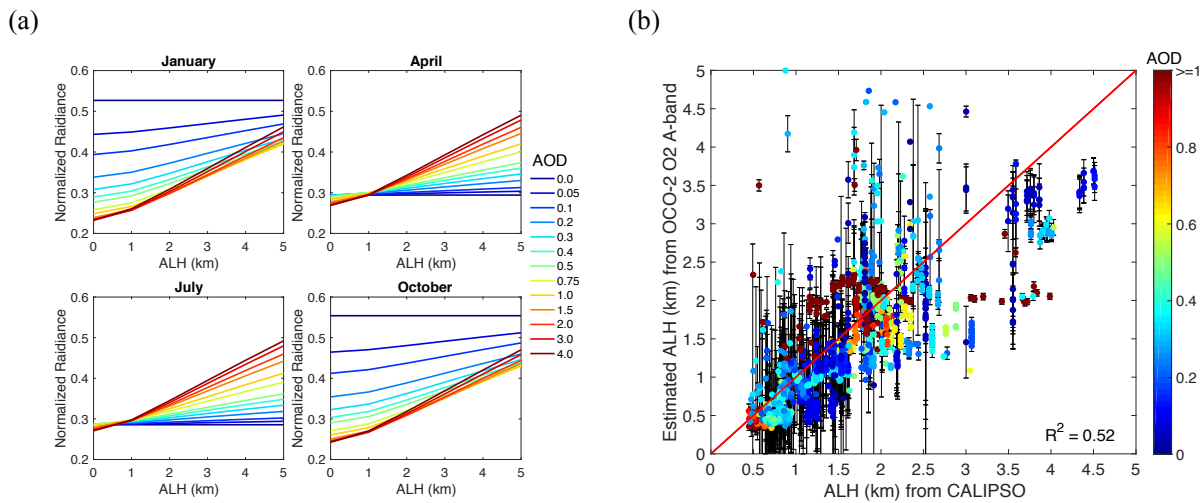

(a)

(b)

**Figure 9**. (a) Examples of ALH look up tables for four months. These LUTs are selected from the LUT database (as shown in **Figure A6**) according to the SZA, wind speed, and atmospheric pressure of the sounding; (b) Scatter plot of ALH estimates from OCO-2 $O_2$ A-band measurements and CALIPSO, color-coded by AOD, over the western Sahara coast. The RMSE is 0.68 km. About 84% of the measurements fall into the calculated range and generate valid ALH retrievals. Some of the error bars are outside the figure box. Error bars for data points with uncertainties larger than 2 km are not shown (which accounts for 4% of the data).

## 5. Discussion

### 5.1 Challenges in applying spectral sorting approach to measurements over land

There are two key challenges when trying to apply our spectral sorting approach to satellite measurements of the $O_2$ A-band over land. The first is the determination of surface reflectance. The surface reflectance in the $O_2$ A-band over land has a large range of variability for different land surface types (**Moody et al., 2005**). Especially over urban regions, the main source of anthropogenic aerosol emissions, the complex landscape makes the BRDF very challenging to model. Any error in surface BRDF estimates will propagate to the retrieved AOD and ALH. One solution is to make use of long-term measurements of surface BRDF from other satellites such as MODIS (**Schaaf et al., 2002**). The second challenge is the determination of aerosol optical properties, including single scattering albedo and phase function, which are associated with aerosol composition, size distribution, and shape distribution. Errors in our knowledge of aerosol composition and microphysical properties translates into errors in estimates of AOD and ALH. In addition, if the aerosols are absorbing, e.g. wildfire plumes, then the retrieval of ALH will be highly uncertain since the scattering signal is weak. In the v8 OCO-2 FP retrieval algorithm, each sounding includes scattering by thin liquid water and water ice clouds, and the two dominant aerosol types pre-determined from the MERRA reanalysis dataset. There are several other emerging techniques that may provide further constraints for profiling aerosols over land, such as polarization (**Xu et al., 2017**) and multi-angle observations (**Diner et al., 2018**). The AirMSPI instrument (**Xu et al., 2017**) combines polarimetric and hyperspectral oxygen absorption measurements to simultaneously retrieve aerosol vertical structure, aerosol microphysical properties and surface reflection. As shown in **Ding et al. (2016)**, polarization measurements may help improve the sensitivity of detecting aerosols over bright surfaces.

### 5.2. Cloud Contamination

The method developed in this study focuses on clear soundings without cloud contamination. The cloudless OCO-2 sounding tracks were selected with the aid of CALIPSO data. In an operational environment, however, the clouds in the field of view should be identified using the OCO-2 data. For example, the cloud-screening algorithm developed by Taylor et al. (2016) is shown to be highly effective when compared with MODIS cloud data. This method can be combined with the spectral sorting approach in an operational environment. Over ocean, thick clouds, which lead to strong enhancement in continuum radiance can be easily identified (Taylor et al., 2016) since the ocean surface is relatively dark within the $O_2$ A-band. However, the identification of the contamination by thin liquid water or water ice clouds may be challenging. The contamination will lead to an overestimation of AOD and a larger bias in ALH. A detailed analysis of the impact of possible clouds on the retrieval of AOD and ALH is beyond the scope of this study. Taylor et al. (2016) show that thin cirrus clouds are relatively easy to detect in $O_2$ A-band observations because they produce large changes in the optical path length. For the current method, these high cirrus clouds will lead to anomalous overestimation of ALH, which can be identified statistically when compared with the aerosol climatology data. Taylor et al. (2016) find that low-altitude clouds are much more challenging to detect using $O_2$ A-band observations alone because they produce small variations in the optical path lengths. They find that these clouds can be effectively screened using ratios of $CO_2$ and $H_2O$ column abundances retrieved in strongly and weakly absorbing spectral regions. It may be possible to exploit a similar approach with the spectral sorting methods developed here.

## 6. Conclusions

We describe a spectral sorting approach for constraining aerosol optical depth and vertical structure using $O_2$ A-band measurements from OCO-2. The effectiveness of the approach is demonstrated by application to dusty soundings over the western Sahara coast and comparison with co-located lidar measurements from CALIPSO CALIOP. Using the OCO-2 forward model to emulate OCO-2 measurements, we find that the observed $O_2$ A-band hyperspectral measurements have high sensitivity to the aerosol vertical structure. Retrieved estimates of AOD and ALH based on a LUT technique show good agreement with CALIPSO measurements, with correlation coefficients of 0.65 and 0.53, respectively. Compared to conventional fitting schemes, the spectral sorting technique developed for OCO-2 has two strengths: first, spectral channels with high sensitivity to AOD and ALH can be identified; second, different AOD and ALH scenarios show distinct signatures in the sorted radiance spectra and the associated information can be obtained in a straightforward manner.

In the future, we will evaluate this spectral sorting approach over land regions, especially over urban emission plumes where large amounts of aerosols and trace gases are co-emitted. Based on the spectral sorting proposed in this study, a "divide and conquer" strategy may be worth considering for future retrieval algorithm improvement to account for aerosol scattering effects. Instead of performing a simultaneous retrieval of aerosols, trace gases and many other geophysical variables, the retrieval algorithm could first extract information on aerosol vertical structure from $O_2$ A-band measurements using the proposed spectral sorting approach. Then, the retrieved AOD and ALH can be utilized by the FP retrieval algorithm for trace gas retrieval.

## Acknowledgement

The OCO-2 Forward model is available at https://github.com/nasa/RtRetrievalFramework. The L1bSc OCO-2 radiances are available online from the NASA Goddard GES DISC at https://disc. gsfc.nasa.gov/datacollection/OCO2_ L1B_Science_7.html. MERRAero monthly 3-hour averaged dust column density data can be downloaded from (https://portal.nccs.nasa.gov/cgi-lats4d/webform.cgi?&i=GEOS-5/MERRAero/monthly/tavg3hr_2d_aer_Nx). S. C. acknowledges support from the SURF program at the California Institute of Technology and from the National University of Singapore. V. N. acknowledges support from the NASA Earth Science US Participating Investigator program (solicitation NNH16ZDA001N-ESUSPI). F. X. acknowledges support from the NASA Remote Sensing Theory program under grant 14-RST14-0100. We are also thankful for support from the Jet Propulsion Laboratory Research and Technology Development Program.

**Declaration of conflicts of interest:**

none.

**APPENDIX A1: Histograms of AOD and ALH from CALIPSO**

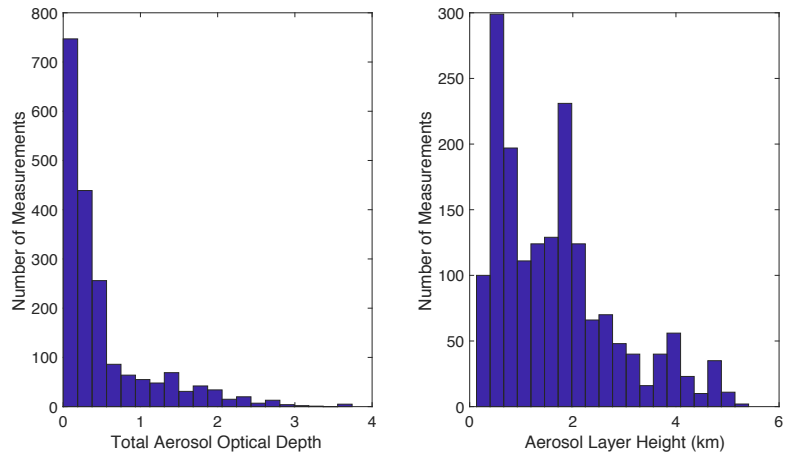

**Figure A1.** Histograms of AOD at 0.76 um and ALH calculated from CALIPSO measurements over the western Sahara coast. There are a total of 45 OCO-2 sounding tracks from 2014 to 2018. After excluding tracks contaminated by clouds and not collocated with CALIPSO, there are 27 sounding tracks available for this study (as listed in **Appendix A2**). The AOD is interpolated based on the Ångström exponent law as described in **Appendix A4**. The algorithm for calculating ALH is described in **Appendix A3**. The average AOD and ALH values are 0.33 and 1.60 km, respectively.

# APPENDIX A2: List of all available OCO-2 sounding tracks used in this study

**Table A2**. Information about the 27 OCO-2 sounding tracks used in this study, including the sensing date and the orbit number and mode. The OCO-2 data on these days can be accessed from the NASA Earth Data Portal (https://search.earthdata.nasa.gov/).

| Track # | Date | Orbit & Mode | Track # | Date | Orbit & Mode |
|---|---|---|---|---|---|
| 1 | 141014 | 01519a | 15 | 161003 | 12004a |
| 2 | 150219 | 03383a | 16 | 161104 | 12470a |
| 3 | 150323 | 03849a | 17 | 161206 | 12936a |
| 4 | 150510 | 04548a | 18 | 170208 | 13868a |
| 5 | 150713 | 05480a | 19 | 170312 | 14334a |
| 6 | 150814 | 05946a | 20 | 170515 | 15266a |
| 7 | 150915 | 06412a | 21 | 170616 | 15732a |
| 8 | 151220 | 07810a | 22 | 171022 | 17596a |
| 9 | 160121 | 08276a | 23 | 180126 | 18994a |
| 10 | 160325 | 09208a | 24 | 180227 | 19460a |
| 11 | 160426 | 09674a | 25 | 180331 | 19926a |
| 12 | 160528 | 10140a | 26 | 180502 | 20392a |
| 13 | 160629 | 10606a | 27 | 180907 | 22256a |
| 14 | 160901 | 11538a | | | |

## APPENDIX A3: Interpolation of AOD using Ångström exponent law

CALIPSO makes AOD measurements at 532 and 1064 nm. The AOD value in the $O_2$ A-band at 0.76 µm can be estimated using the Ångström exponent law (Seinfeld and Pandis, 2006):

$$\frac{\tau}{\tau_0} = \left(\frac{\lambda}{\lambda_0}\right)^{-k}$$

(A4)

where $\lambda$ is the wavelength and $\tau$ is the corresponding AOD to be estimated; $\lambda_0$ is the reference wavelength and $\tau_0$ is the corresponding AOD from CALIPSO; $k$ is the Ångström exponent, which can be calculated from CALIPSO AOD measurements at its two wavelengths.

## APPENDIX A4: Calculation of ALH from CALIPSO aerosol profile

The ALH is defined as the center of mass of aerosol scatterers. We calculate ALH using the widely used weighted mean method as described in Xu et al. (2017) and Koffi et al. (2012):

$$ALH = \frac{\sum_{i=1}^{n} \beta_i \cdot Z_i}{\sum_{i=1}^{n} \beta_i} \tag{A2}$$

where $\beta_i$ is the CALIPSO aerosol extinction coefficient and $Z_i$ is the corresponding height at level $i$. The derived ALH is different from the peak height ($x_a$) when $x_a$ is small. By definition, the ALH will be equal to or larger than $x_a$ for a Gaussian aerosol profile. The relationship between them is shown in **Figure A4**. In addition, $x_a$ is measured by the relative pressure, P/P$_{surf}$, while the ALH is the geometric length in units of kilometers. The conversion between height (h) in kilometers and relative pressure (P/P$_{surf}$) is based on the following empirical equation:

$$\frac{P(h)}{Psurf} = \exp{(h/H)} \tag{A3}$$

where H is the atmospheric scale height, which is about 9 km over the western Sahara coast region. Based on **Equation (A3),** the atmospheric scale height is calculated from the pressure profile and the corresponding heights for each pressure level selected from the GEOS FP-IT reanalysis (**Lucchesi et al., 2013**).

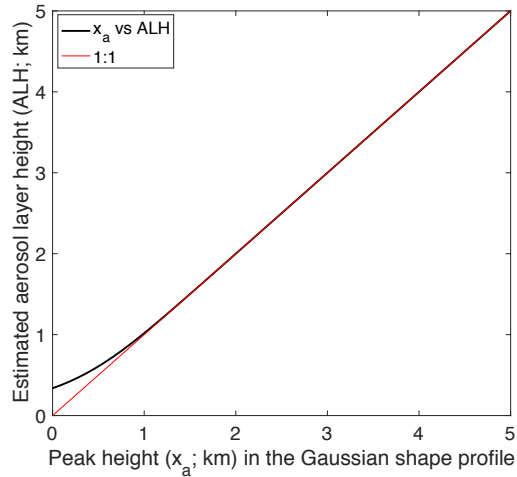

**Figure A4**. The correlation between the peak height ($x_a$) in a Gaussian profile and the aerosol layer height (ALH) estimated from **Equation (A2)**. The one-to-one line is also indicated.

**APPENDIX A5: Single scattering albedo and phase function for dust aerosol**

The single scattering albedo and phase function of dust aerosol used in the OCO-2 forward model is adopted from MERRA climatology (2009-2010; **Rienecker et al., 2011**). The dust single scattering albedo is 0.93, and its phase function is shown in **Figure A5**.

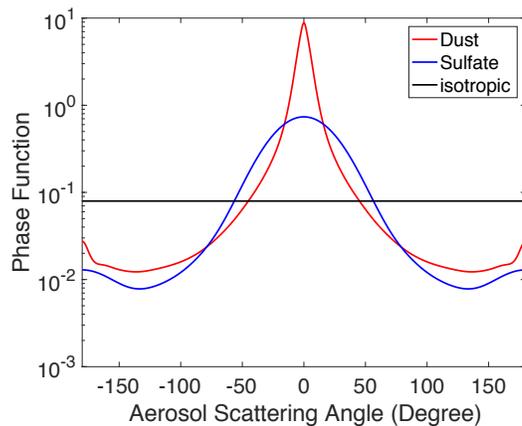

**Figure A5**. The phase function of dust aerosol for $O_2$ A-band, derived from MERRA reanalysis data, and used in the OCO-2 forward model. As a comparison, the phase functions for sulfate and isotropic scattering are also plotted.

## APPENDIX A6: Examples of lookup tables for AOD and ALH retrievals

Due to changes in observation geometry and atmospheric conditions on a daily basis, the LUTs were built as a function of varying parameters, including solar zenith angle (SZA), wind speed, atmospheric pressure, AOD, and ALH. Examples of the LUTs are shown in **Appendix A6**. The enhancement of radiance from increased AOD (panel (a)) and enhancement of the normalized radiance from increased ALH (panel (b)) can be clearly observed.

(a)

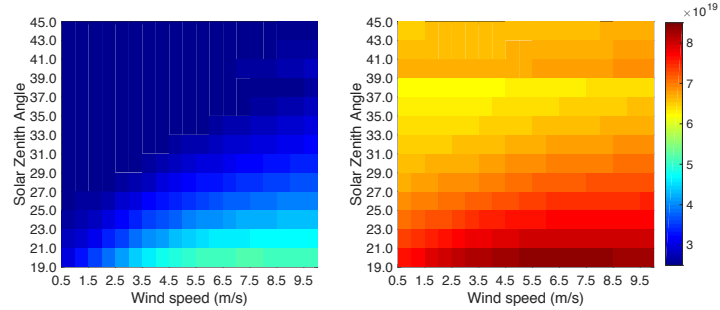

(b)

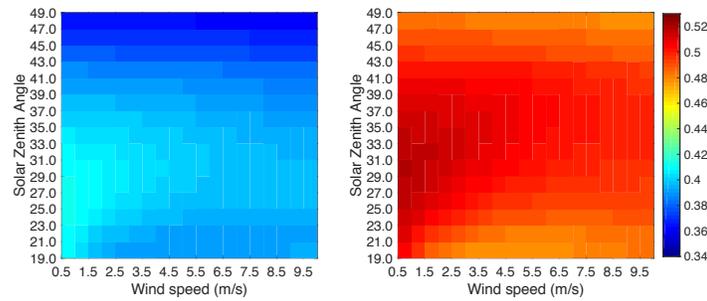

**Figure A6**. Examples of look-up tables (LUTs) as a function of solar zenith angle and wind speed. (a) LUTs for AODs of 0.3 (left) and 1.0 (right). The data for each grid is the simulated $O_2$ A-band continuum radiance in unit of (Photons sec$^{-1}$ m$^{-2}$ sr$^{-1}$ um$^{-1}$). The ALHs are 1.0 km for both cases; (b) LUTs for ALHs of 1.0 km (left) and 3.0 km (right). The AODs are 0.3 for both cases. The data for each grid is the normalized radiance in $O_2$ A-band. See **Section 3.3** for details.

# APPENDIX A7: Lookup Table Uncertainty

The uncertainty of LUT is related to uncertainties in the OCO-2 RT forward model and the associated model inputs. While it can be very complex to quantify LUT uncertainty analytically, several variables are important, including wind speed, atmospheric pressure, and aerosol optical property. The details of the uncertainty quantification are described in **Section 4**.

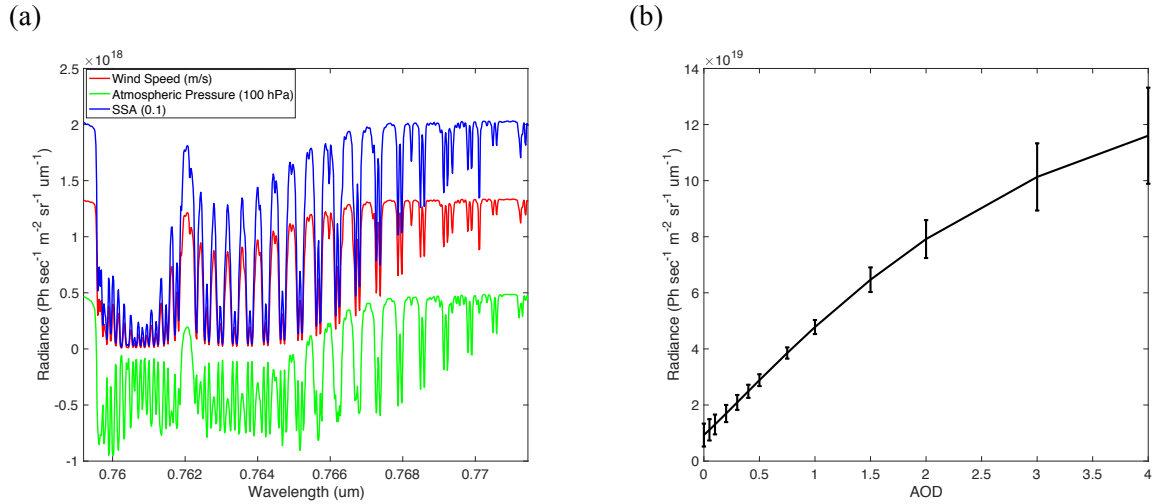

**Figure A7**. Example of LUT uncertainty. (a) The Jacobians of $O_2$ A-band radiance to wind speed atmospheric pressure, and aerosol single scattering albedo, are estimated by finite difference using OCO-2 forward model. These examples are adopted from case when AOD=0.3 and ALH=1.6km on 14 October 2014. The wind speed is 5.5m/s and the SZA is 35.6degrees; (2) Example of the LUT uncertainty for the LUT on 14 October 2014. The error bars indicate the uncertainty calculated using **Equation (2)**.

## APPENDIX A8: Difference between LUT and OCO-2 measurements

To assess the performance of the OCO-2 forward model, we compare the continuum level radiance between OCO-2 measurements and the RT forward model when AOD is small (less than 0.3). The results (**Figure A8**) show that the one sigma (standard deviation) difference is about 21%. This difference comes from (1) the uncertainty of the forward model and (2) the colocation error between CALIPSO and OCO-2. CALIPSO data has a 5-km resolution while OCO-2's footprint size is about 1 km. This mismatch may also explain some of the uncertainty between our retrievals of AOD and ALH (**Figure 8** and **Figure 9**) and the corresponding CALIPSO data.

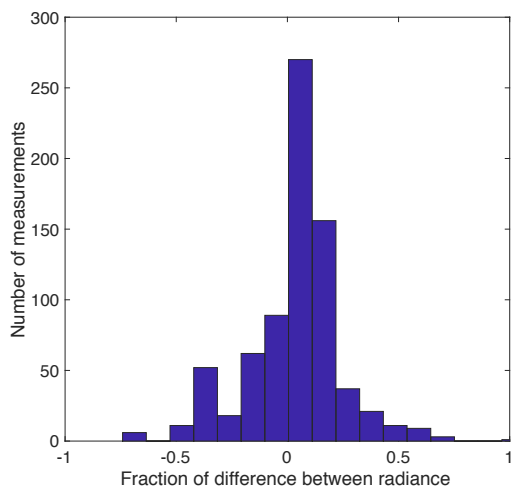

**Figure A8**. Fractional difference in continuum level radiance between OCO-2 $O_2$ A-band measurements and OCO-2 RT forward model calculations when AOD is less than 0.3. The standard deviation of the difference is 0.21.